\documentclass[journal]{IEEEtran}

\usepackage{url}
\usepackage{bs_macros}
\usepackage{caption}
\usepackage{float}
\usepackage[labelformat=simple]{subcaption}

\usepackage{balance}

\graphicspath{{./}}

\theoremstyle{definition}

\newtheorem{remark}{Remark}

\newtheorem{example}{Example}

\makeatletter
\patchcmd{\@maketitle}
{\addvspace{0.5\baselineskip}\egroup}
{\addvspace{-1.25\baselineskip}\egroup}
{}
{}
\makeatother

\begin{document}

\title{Unitary Shift Operators on a Graph}

\author{
	
	Bruno Scalzo Dees, \IEEEmembership{Student Member, IEEE}, Ljubi$\check{\text{s}}$a Stankovi\'c, \IEEEmembership{Fellow, IEEE}, Milo$\check{\text{s}}$ Dakovi\'c, \IEEEmembership{Member, IEEE}, Anthony G. Constantinides, \IEEEmembership{Life Fellow, IEEE}, Danilo P. Mandic, \IEEEmembership{Fellow, IEEE} 
	
	\thanks{B. Scalzo Dees, A. G. Constantinides and D. P. Mandic are with the Department of Electrical and Electronic Engineering, Imperial College London, London SW7 2AZ, U.K., e-mail: \{bruno.scalzo-dees12, a.constantinides, d.mandic\}@imperial.ac.uk.}
	\thanks{L. Stankovi\'c and M. Dakovi\'c are with the with the Faculty of Electrical Engineering, University of Montenegro, D$\check{\text{z}}$ord$\check{\text{z}}$a Va$\check{\text{s}}$ingtona bb, 81000 Podgorica, Montenegro, e-mail: \{ljubisa, milos\}@ucg.ac.me.}
	
}


\maketitle

\begin{abstract}
	A unitary shift operator (GSO) for signals on a graph is introduced, which exhibits the desired property of energy preservation over both backward and forward graph shifts. For rigour, the graph differential operator is also derived in an analytical form. The commutativity relation of the shift operator with the Fourier transform is next explored in conjunction with the proposed GSO to introduce a graph discrete Fourier transform (GDFT) which, unlike existing approaches, ensures the orthogonality of GDFT bases and admits a natural frequency-domain interpretation. The proposed GDFT is shown to allow for a coherent definition of the graph discrete Hilbert transform (GDHT) and the graph analytic signal. The advantages of the proposed GSO are demonstrated through illustrative examples.
\end{abstract}


\IEEEpeerreviewmaketitle


\vspace{-0.2cm}

\section{Introduction}

The rapidly expanding field of Graph Signal Processing \cite{Sandryhaila2013,Shuman2013,Ortega2018,Stankovic2019_1} requires a rigorous definition of a \textit{graph shift operator}, which should ideally incorporate domain-specific knowledge of the graph topology as a means to introduce the graph counterparts of many classical signal processing techniques. While the signal shift is straightforwardly defined on the discrete-time axis, its definition on irregular graph domains is not obvious due to the rich underlying connectivity structure. 

Existing GSOs typically take the convenient form of the graph adjacency or Laplacian matrices which are not isometric operators, however, for rigour it is desirable, or even necessary, to preserve the signal energy ($L_{2}$-norm) over shifts. Consequently, several isometric shift operators have been recently proposed which satisfy the energy-preserving property \cite{Girault2015,Girault2015_2,Gavili2017}, whereby the eigenvalues of the adjacency or Laplacian matrix are cast onto a unit circle, thus preserving in this way the isometry property. However, the corresponding GDFT harmonics are not orthogonal for a general graph structure. 

To this end, we set out to revisit the definition of unitary GSOs. This is achieved by drawing inspiration from the classical definition of the \textit{unitary} shift operator acting on Hilbert spaces \cite{Fillmore1974}. Its continuous counterpart, the \textit{graph differential operator}, is also derived. Owing to the isometric nature of the proposed GSO, we show that its eigen-analysis provides a new basis for the graph discrete Fourier transform which exhibits a frequency-domain interpretation as in classical Fourier analysis. As a result, rigorous definitions of the graph discrete Hilbert transform and the graph analytic signal are provided based on the proposed GSO. Practical utility of the proposed class of GSOs, and its underlying GDFT functionality, is demonstrated through intuitive examples.



\section{Unitary Graph Shift Operators}

We follow the notation employed in \cite{Stankovic2019_1} whereby a graph, denoted by $\mathcal{G} = \{\mathcal{V},\mathcal{E}\}$, is defined as a set of $N$ vertices, $\mathcal{V} = \{1,2,...,N\}$, which are connected by a set of edges, $\mathcal{E} \subset \mathcal{V} \times \mathcal{V}$. The existence of an edge between vertices $m$ and $n$ is designated by $(m, n) \in \mathcal{E}$. The graph connectivity of an $N$-vertex graph can be formally represented by the \textit{adjacency matrix}, $\A \in \domR^{N \times N}$, whereby the vertex connectivity structure is described by \vspace{-0.3cm}
\begin{align}
	A_{mn} = \begin{cases}
		1, & (m,n) \in \mathcal{E},\\
		0, & (m,n) \notin \mathcal{E}.
	\end{cases} \label{eq:adjacency}
\end{align}


\subsection{Properties of shifted signals}

Consider a graph signal, $\x \in \domR^{N}$, for which $x(n)$ is the observed sample at a vertex $n \in \mathcal{V}$. Topologically, a \textit{necessary} property of the \textit{backward shift} on a graph is that the signal sample, $x(n)$, must in some respect move from its original vertex, $n$, to vertices, $m$, within its nearest neighborhood modelled by $\A$ in (\ref{eq:adjacency}), with $(m,n)\in \mathcal{E}$ (see Figure \ref{fig:1}). In its elementary form, the backward shifted signal, $\y_{b} \in \domR^{N}$, can thus be defined using the graph adjacency matrix, $\A$, as $\y_{b} = \A\x$ \cite{Sandryhaila2013}.

\begin{remark} \label{remark:colspace}
	The adjacency matrix, $\A$ in (\ref{eq:adjacency}), provides the minimal information required to fully reflect the connectivity structure arising from the graph topology, and therefore to define the most elementary graph shift. Therefore, a sufficient condition to define the backward graph shift is that the shifted signal, $\y_{b}$, lies in the \textit{column space} of $\A$.
\end{remark}

Similarly, the \textit{forward shift} on a graph can be defined as a movement of the signal sample, $x(n)$, from its original vertex, $n$, to vertices, $m$, within its neighborhood for which $(n,m) \in \mathcal{E}$. The forward shifted signal, denoted by $\y_{f} \in \domR^{N}$, can also be defined using the graph adjacency matrix, $\A$, as $\y_{f} = \A^{\Trans}\x$.

\begin{remark}
	Following Remark \ref{remark:colspace}, a sufficient condition to define the forward graph shift is that the shifted signal, $\y_{f}$, lies in the \textit{row space} of $\A$. For the visualisation, see \mbox{Figure \ref{fig:1}}.
\end{remark}

It is important to notice that for a general adjacency matrix, $\A$, the signal energy is not necessarily preserved over shifts, that is, $\|\A\x\| \neq \|\x\|$ or $\|\A^{\Trans}\x\| \neq \|\x\|$. However, it is often desirable, or even necessary, that a graph shift does not decrease or increase signal energy. It would also be advantageous to have a shift operator whereby the forward shift represents the \textit{inverse} of the backward shift -- not the case with $\A$. One such isometric GSO is introduced in the following.

\pagebreak

\subsection{Unitary graph shift operator}

Following on the adjacency matrix based definition of the shift operator, we shall next consider the task of determining a GSO, denoted by $\S \in \domR^{N \times N}$, which exhibits the properties of shift operators acting on Hilbert spaces \cite{Fillmore1974}, given by: \vspace{-0.05cm}
\begin{enumerate}[label=\roman*),leftmargin=5mm]
	\item The row and column spaces of $\S$ and $\A$ coincide;
	\item The forward GSO is the inverse of the backward GSO;
	\item The energy of any graph signal, $\x \in \domR^{N}$, is preserved over both backward and forward shifts, that is \vspace{-0.1cm}
	\begin{align}
		\|\S\x\|=\|\S^{-1}\x\|=\|\x\| \label{eq:isometry}
	\end{align} \vspace{-0.5cm} \\
	In other words, $\S$ is an isometric mapping.
\end{enumerate}

\vspace{-0.2cm}

\begin{remark}
	For the condition in (\ref{eq:isometry}) to hold for any graph signal, $\x \in \domR^{N}$, the matrix $\S$ must be \textit{unitary}, whereby the conditions $\S^{-1}=\S^{\Trans}$ and $\S^{\Trans}\S = \S\S^{\Trans} = \I$ should be satisfied.
\end{remark}

\vspace{-0.15cm}

The task of defining an isometric GSO can now be formalised as that of determining the unitary matrix, $\S$, which is closest to $\A$ in a Hilbert space, that is, $\S$ is required to exhibit the \textit{maximum inner product} with $\A$, that is \vspace{-0.1cm}
\begin{align}
	\max_{\S} \quad & \inner{\S}{\A} = \sum_{m=1}^{N}\sum_{n=1}^{N}S_{mn}A_{mn} = \tr{ \S\A^{\Trans} } \notag \\
	\textnormal{s.t.} \quad &  \S^{\Trans}\S = \S\S^{\Trans} = \I \label{eq:optimization}
\end{align}
This can be achieved analytically by evaluating the singular value decomposition of $\A$, given by \vspace{-0cm}
\begin{align}
	\A = \U \boldSigma  \V^{\Trans} \label{eq:SVD}
\end{align} \vspace{-0.5cm}  \\
where $\U, \V \in \domR^{N \times N}$ are respectively the left and right matrix of singular vectors, and $\boldSigma \in \domR^{N \times N}$ is the diagonal matrix of singular values.  From (\ref{eq:optimization}) and (\ref{eq:SVD}), the backward shift operator can be expressed as \vspace{-0.2cm}
\begin{align}
	\S = \U\Q\V^{\Trans} \label{eq:GSO_sym_orth}
\end{align} \vspace{-0.3cm}
where \vspace{-0.1cm}
\begin{align}
	\Q = \left[\begin{array}{cc}
		\I_{(N-1) \times (N-1)} & \0_{(N-1) \times 1} \\
		\0_{1 \times (N-1)} & \det(\U\V^{\Trans})
	\end{array}\right]
\end{align}
ensures that $\det(\S)=1$, so as to produce a \textit{proper rotation} matrix. The maximum inner product with $\A$ is therefore equal to $\inner{\S}{\A} = \tr{ \boldSigma }$. 

\vspace{-0.1cm}

\begin{remark}
Since $\U$, $\Q$ and $\V$ are \textit{unitary} matrices, $\S$ preserves signal energy over both backward and forward shifts. Furthermore, $\U$ and $\V$ provide respectively the orthonormal bases for the column and row spaces of $\A$. Therefore, the row and column spaces of $\S$ coincide with those of $\A$.
\end{remark}
	
\vspace{-0.3cm}
	
\begin{remark}
	The matrix $\S = \U\Q\V^{\Trans}$ is called the \textit{symmetric orthogonalization} of the matrix $\A$, and is unique \cite{Lowdin1950,Lowdin1970}. The solution is also closely related to the orthogonal Procrustes problem \cite{Schonemann1966} and the associated Kabsch algorithm \cite{Kabsch1976}. An important feature is that among all possible orthogonalizations of $\A$, the symmetric orthogonalization ensures that $\|\S-\A\|$ is minimised, or equivalently, $\inner{\S}{\A}$ is maximised \cite{Pratt1956}. Therefore, the column and row spaces of $\S$ are the closest in the Hilbert space to those of $\A$. 
\end{remark}

\vspace{-0.3cm}


\begin{remark}
	For an undirected graph, whereby $\A=\A^{\Trans}$, the proposed shift operator is both symmetric, $\S = \S^{\Trans}$, and unitary, $\S = \S^{-1}$, which means that the forward and backward shifts are equivalent.
\end{remark}

\pagebreak






%

\begin{figure}[ht]
	\centering
	\begin{subfigure}[t]{0.24\textwidth}
		\centering
		\includegraphics[width=0.8\textwidth, trim={7.2cm 22cm 7.4cm 1.9cm}, clip]{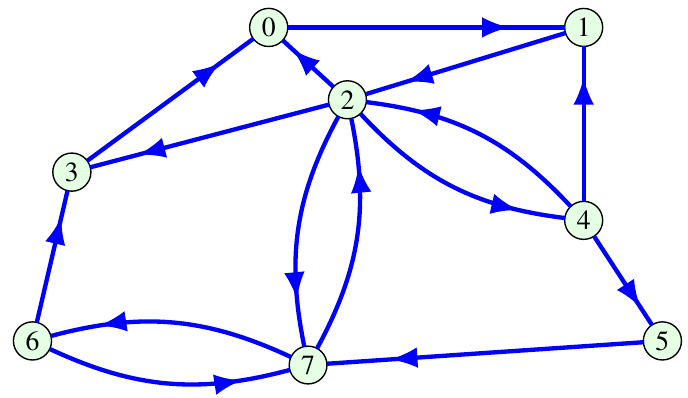} 
		\caption[]%
		{{\small Directed graph structure.}}    
		\label{fig:structure}
	\end{subfigure}
	\hfill
	\begin{subfigure}[t]{0.24\textwidth}
		\centering
		\includegraphics[width=0.9\textwidth, trim={4cm 12cm 4cm 11cm}, clip]{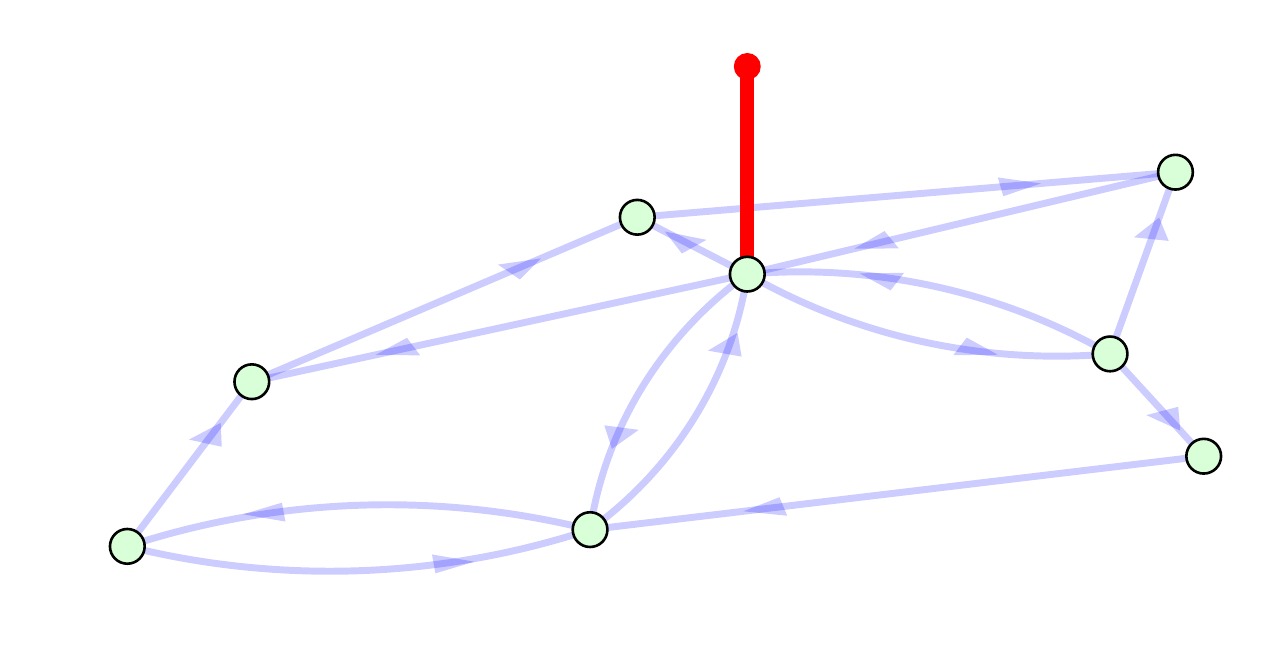} 
		\caption[]%
		{{\small Graph signal, $\x$.}}    
		\label{fig:orig}
	\end{subfigure}
	\vspace{-0.4cm}
	\vskip\baselineskip
	\centering
	\begin{subfigure}[t]{0.24\textwidth}
		\centering
		\includegraphics[width=0.9\textwidth, trim={4cm 12.8cm 4cm 10.5cm}, clip]{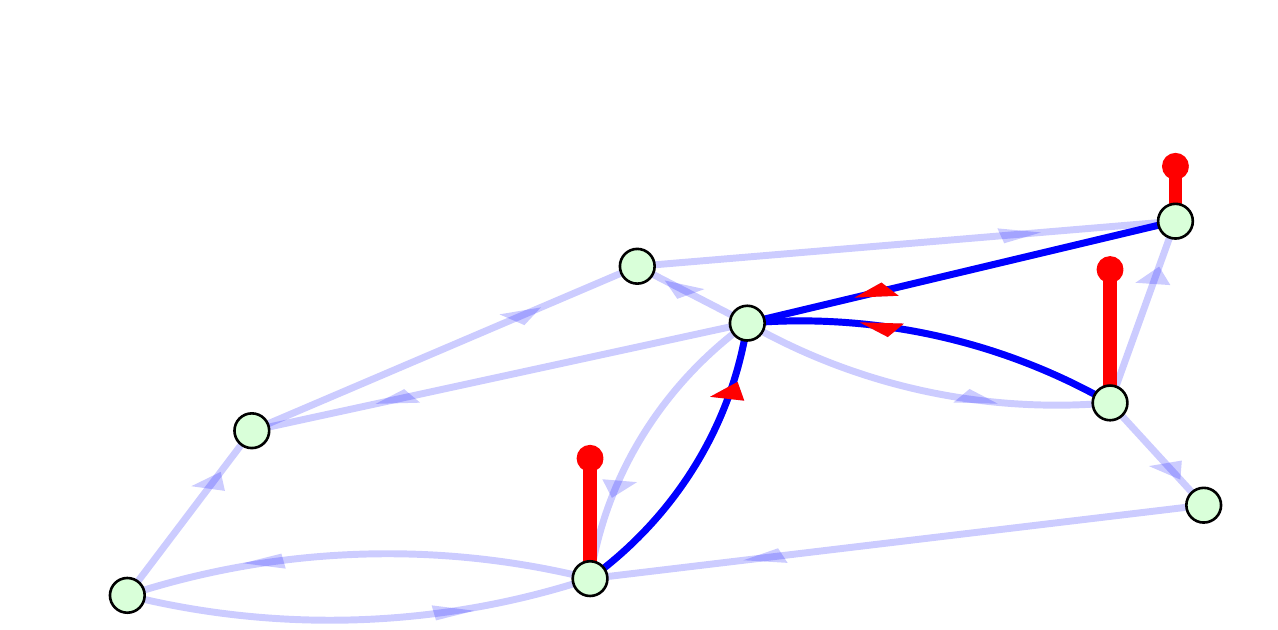} 
		\caption[]%
		{{\small Backward shifted signal, $\S\x$.}}    
		\label{fig:Sbshift}
	\end{subfigure}
	\hfill
	\begin{subfigure}[t]{0.24\textwidth}  
		\centering 
		\includegraphics[width=0.9\textwidth, trim={4cm 12.8cm 4cm 10.5cm}, clip]{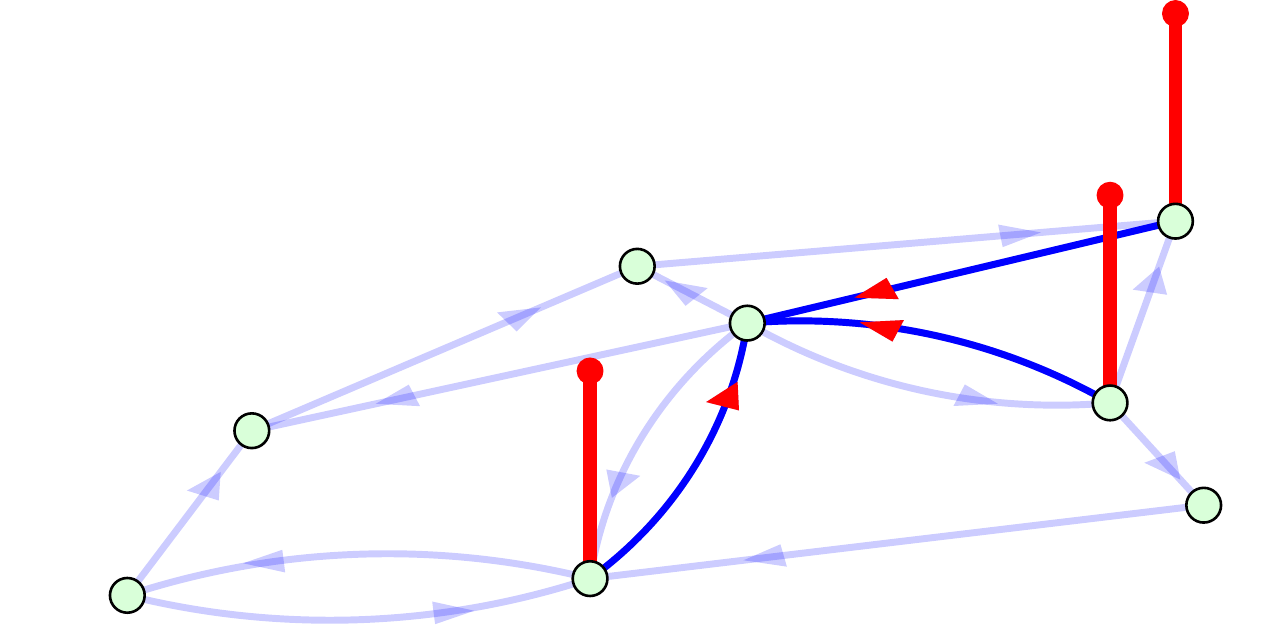} 
		\caption[]%
		{{\small Backward shifted signal, $\A\x$.}}    
		\label{fig:Abshift}
	\end{subfigure}
	\vspace{-0.4cm}
	\vskip\baselineskip
	\begin{subfigure}[t]{0.24\textwidth}   
		\centering 
		\includegraphics[width=0.9\textwidth, trim={4cm 12.8cm 4cm 10.9cm}, clip]{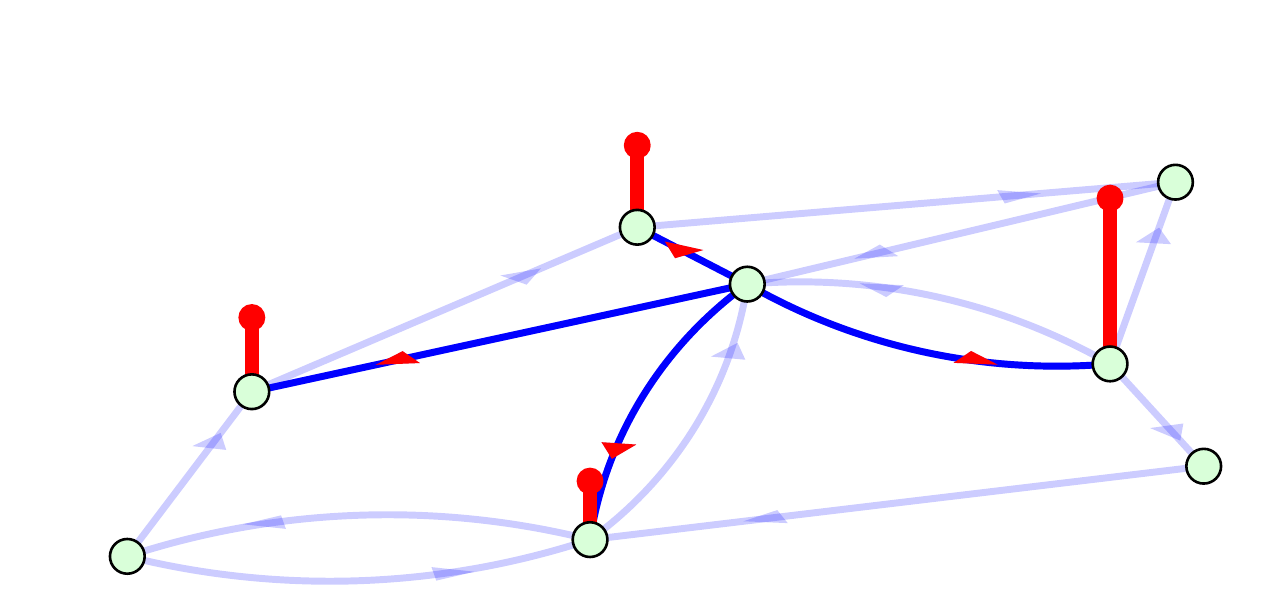} 
		\caption[]%
		{{\small Forward shifted signal, $\S^{\Trans}\x$.}}    
		\label{fig:Sfshift}
	\end{subfigure}
	\hfill
	\begin{subfigure}[t]{0.24\textwidth}   
		\centering 
		\includegraphics[width=0.9\textwidth, trim={4cm 12.8cm 4cm 10.9cm}, clip]{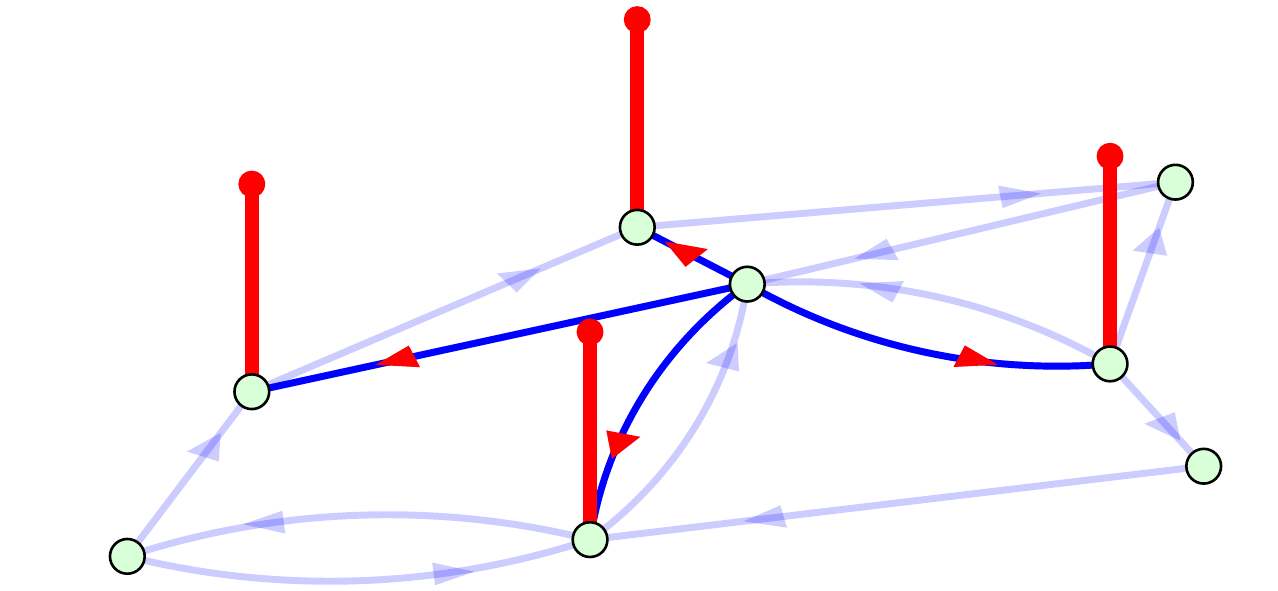} 
		\caption[]%
		{{\small Forward shifted signal, $\A^{\Trans}\x$.}}    
		\label{fig:Afshift}
	\end{subfigure}
	\caption[]
	{\small Graph signal shifts by the proposed unitary shift operator, $\S$, evaluated on a directed graph. (a) Directed graph structure. (b) A simple graph signal, $\x$. (c) Backward shifted version of $\x$, given by $\S\x$. (d) Backward shifted version, $\A\x$. (e) Forward shifted version, $\S^{\Trans}\x$. (f) Forward shifted version, $\A^{\Trans}\x$. The red arrows indicate the movement of the pulse at vertex $n=2$ towards vertices connected with (i) outgoing blue arrows in (a) for a forward shift and (ii) to vertices connected with incoming blue arrows for a backward shift.} 
	\label{fig:1}
\end{figure}

\begin{example}
	Consider the directed graph in Figure 1(a) which has $N=8$ vertices in the set $\mathcal{V}=\{0,1,2,3,4,5,6,7\}$. The corresponding adjacency matrix, $\A$, is given by \vspace{-0.4cm}
	\begin{gather}
		\A  = 
		\arraycolsep=3pt
		\def\arraystretch{0.9}
		\begin{array}{cr}
		& \\
		{
			\color{blue}
			\begin{matrix}
			\text{\footnotesize 0}\\
			\text{\footnotesize 1}\\
			\text{\footnotesize 2}\\
			\text{\footnotesize 3}\\
			\text{\footnotesize 4}\\
			\text{\footnotesize 5}\\
			\text{\footnotesize 6}\\
			\text{\footnotesize 7}\\
			\end{matrix}
		} & 
		\left[
		\begin{array}{cccccccc}
		0  &  {\color{red}1}  &  0  &  0  &  0  &  0  &  0  & 0 \\
		0  &  0  &  {\color{red}1}  &  0  &  0  &  0  &  0  & 0  \\
		{\color{red}1}  &  0  &  0  &  {\color{red}1}  &  {\color{red}1}  &  0  &  0  & {\color{red}1}  \\
		{\color{red}1}  &  0  &  0  &  0  &  0  &  0  &  0  & 0  \\
		0  &  {\color{red}1}  &  {\color{red}1}  &  0  &  0  &  {\color{red}1}  &  0  & 0  \\
		0  &  0  &  0  &  0  &  0  &  0  &  0  & {\color{red}1}  \\
		0  &  0  &  0  &  {\color{red}1}  &  0  &  0  &  0  & {\color{red}1}  \\
		0  &  0  &  {\color{red}1}  &  0  &  0  &  0  &  {\color{red}1}  & 0 
		\end{array}
		\right]
		\end{array} \label{eq:Adjacency}
	\end{gather}
	The backward and forward shifted versions of the signal in Figure \ref{fig:orig} were evaluated using both the elementary shift matrix, $\A$, and the proposed isometric GSO, $\S$ in (\ref{eq:GSO_sym_orth}), and are illustrated in Figure \ref{fig:1}. Notice that, as desired, the signal energy was preserved when employing the unitary GSO, $\S$, while the energy of the signals shifted through $\A$ increased.
\end{example}

\vspace{-0.5cm}

\vspace{-0.5cm}

\begin{remark}
Like in standard linear shift-invariant systems, a system on a graph can be implemented as a linear combination of a graph signal, $\x$, and its graph shifted versions, $\S^{m}\x$. The output signal of an order-$M$ system is then defined as \cite{Sandryhaila2013,Stankovic2019_2} \vspace{-0.1cm}
\begin{align}
	\y = \sum_{m=0}^{M} h_{m} \S^{m}\x \label{eq:system}
\end{align}
where $h_{m}$ are the system coefficients. Owing to the unitary properties of the proposed class of GSOs, systems based on this isometric shift exhibit the desirable \textit{boundedness} property
\begin{align}
	\|\y\|^{2} \leq \sum_{m=0}^{M} |h_{m}|^{2} \|\x\|^{2}
\end{align}
which stems from the triangle inequality employed above.
\end{remark}

\pagebreak

\subsection{Differential operator on a graph}

While the time shift designates a discrete change in a system, the continuous-time counterpart of the shift is the \textit{differential}. A fundamental result which links the discrete and continuous calculus, known as \textit{Stone's theorem} \cite{Stone1930,Stone1932}, provides a direct relation between these operators. This becomes obvious when considering the Taylor series expansion of the shifted signal at a time instant $k$, which has the form
\begin{align}
	\mathcal{S}x(k) & = x(k-1) \notag\\
	& = x(k) - \frac{dx(k)}{dk} + \frac{1}{2!}\frac{d^{2}x(k)}{dk^{2}} - \frac{1}{3!}\frac{d^{3}x(k)}{dk^{3}} + \cdots \notag\\
	& = \left( \sum_{m=0}^{\infty} \frac{(-1)^{m}}{m!} \left(\frac{d}{dk}\right)^{m}   \right) x(k) = \exp\left(-\frac{d}{dk}\right)x(k)
\end{align}
In other words, the shift operator is an exponential of the differential operator. In the graph setting, this relation becomes
\begin{align}
	\S = \exp\left( - \boldnabla \right)
\end{align}
where $\boldnabla \in \domR^{N \times N}$ is the \textit{graph differential operator}. Therefore, the graph differential operator is given by
\begin{align}
	{\boldsymbol \nabla} = - \ln(\S) \label{eq:graph_differential}
\end{align}
and is also referred to as the \textit{infinitesimal generator} of the shift operator, $\S$.

\begin{remark}
	The logarithm of the shift (rotation) matrix, $\S$, exhibits the \textit{skew-symmetry property}, $\boldnabla^{\Trans} = -\boldnabla$. Skew-symmetric matrices form a tangent space of orthogonal matrices, and can be thus thought of as \textit{infinitesimal rotations}.
\end{remark}


\section{Graph Discrete Fourier Transform}

It is well understood that the classical Fourier transform is intrinsically linked to the shift operator. To see this, consider the discrete-time signal, $x(k) \in \domR$, at a time instant, $k$, which exhibits the following Fourier relationship
\begin{align}
	x(k) & \FourierArrow \mathcal{F}\{x\}(\omega)
\end{align}
where $\mathcal{F}\{x\}(\omega) \in \domC$ is the Fourier transform of $x(k)$, and $\omega$ the angular frequency. 

The operation of a shift can then be interpreted through the \textit{time shift property} of the Fourier transform, given by
\begin{align}
	\mathcal{S}x(k) = x(k-1) \FourierArrow e^{-\jmath \omega}\mathcal{F}\{x\}(\omega) \label{eq:shift_theorem}
\end{align}
Owing to the \textit{linearity} property of the Fourier transform, this means that the linear shift operator, $\mathcal{S}$, will also directly apply in the Fourier domain, to yield
\begin{align}
	\mathcal{S}x(k) \FourierArrow \mathcal{S}\mathcal{F}\{x\}(\omega) = e^{-\jmath \omega}\mathcal{F}\{x\}(\omega) \label{eq:diagonalization_shift_theorem}
\end{align} 
From the operator-theoretic perspective, the relation in (\ref{eq:diagonalization_shift_theorem}) can be viewed as the \textit{diagonalization} of the shift operator by the Fourier transform, whereby the eigenfunction, $\mathcal{F}\{x\}(\omega)$, is accompanied by the eigenvalue, $e^{-\jmath \omega}$. In other words, this asserts the classical \textit{commutativity} relationship between the shift operator and the Fourier transform.


\subsection{Graph discrete Fourier transform (GDFT)}


Since the proposed shift, $\S$, is a rotation matrix, its eigenvalues lie on the unit circle and hence take the form $e^{-\jmath \omega}$. Therefore, the commutativity property of the shift operator with the Fourier harmonics in (\ref{eq:diagonalization_shift_theorem}) naturally extends from the discrete-time setting to the irregular graph domain, as follows
\begin{align}
	\S\f_{\omega} = e^{-\jmath \omega}\f_{\omega}
\end{align}
where $\f_{\omega} \in \domC^{N}$ is the graph Fourier harmonic which corresponds to the angular frequency, $\omega$. Equivalently, the proposed GSO, $\S$, admits the following eigenvalue decomposition
\begin{align}
	\S & = \uF \, e^{-\jmath \uboldOmega} \, \uF^{\Her} \label{eq:shift_EVD}
\end{align}
where $\uF \in \domC^{N \times N}$ is the \textit{graph discrete Fourier transform} (GDFT) matrix, and $\uboldOmega \in \domR^{N \times N}$ is a diagonal matrix which contains the frequencies associated to the GDFT harmonics.


\begin{remark}
	The eigenvalues and eigenvectors of $\S$ have a natural frequency-domain interpretation which is in akin to the classical Fourier analysis. The eigenvalue, $e^{-\jmath \omega}$, designates the phase displacement (angular frequency), $\omega$, over a shift, while the associated unit-norm eigenvector, $\f_{\omega}$, defines the \textit{axis of rotation} (harmonic) on a unit-sphere in $\domR^{N}$, whereby the operator $\S$ induces a shift over the unit-sphere.
\end{remark}



Since the shift operator, $\S$, is real-valued, it exhibits a symmetric spectrum. This means that its eigenvalues and eigenvectors are either real-valued or appear in conjugate complex pairs. From (\ref{eq:shift_EVD}), the GDFT matrix takes the form
\begin{align}
	\!\! \uF = \! \left\{ \!\! \begin{array}{l l l}
		\left[
		\arraycolsep=2pt
		\def\arraystretch{1}
		\begin{array}{ccc}
			\f_{0}, & \F, & \F^{\ast}
		\end{array}\right], & \F \in \domC^{N \times \frac{N-1}{2}}, & \text{if $N$ is odd}, \\[1pt]
		\left[
		\arraycolsep=2pt
		\def\arraystretch{1}
		\begin{array}{cc}
			\F, & \F^{\ast}
		\end{array}\right], & \F \in \domC^{N \times \frac{N}{2}}, & \text{if $N$ is even},
	\end{array}\right.
\end{align}
where the column vectors of $\F = \{\f_{\omega}\}_{\omega>0}$ are the GDFT bases associated with the positive frequencies, and $\f_{0}\in \domR^{N}$ is the real-valued bases associated with the DC component ($\omega = 0 $) for the case where $N$ is odd. 

Consequently, the matrix $\uboldOmega$ takes the form
\begin{align}
	\!\! \uboldOmega = \! \left\{ \!\!\! \begin{array}{l l l}
		\left[
		\arraycolsep=1.5pt
		\def\arraystretch{1}
		\begin{array}{ccc}
			0 & \0 & \0  \\
			\0 & \boldOmega & \0 \\
			\0 & \0  & -\boldOmega
		\end{array}\right], & \boldOmega\in\domR^{N \times \frac{N-1}{2}}, & \text{if $N$ is odd}, \\[15pt]
		\left[
		\arraycolsep=1.5pt
		\def\arraystretch{1}
		\begin{array}{cc}
			\boldOmega & \0  \\
			\0 & -\boldOmega 
		\end{array}\right], & \boldOmega\in\domR^{N \times \frac{N}{2}}, & \text{if $N$ is even}, 
	\end{array}\right.
\end{align}
where $\boldOmega$ is a diagonal matrix which contains the positive angular frequencies, $\omega>0$.




Owing to the Hermitian symmetry of $\uF$, the GDFT relationship of a real-valued graph signal, $\x \in \domR^{N}$, is given by
\begin{align}
	\x  =  \uF\,\uX  =  \begin{cases}
		\F\X + \F^{\ast}\X^{\ast} + X(0)\f_{0}, \!\! & \text{if $N$ is odd}, \\[1pt]
		\F\X + \F^{\ast}\X^{\ast}, \!\! & \text{if $N$ is even},
		\end{cases}
\end{align}
where $\uX \in \domC^{N}$ is the GDFT coefficient vector, defined as
\begin{align}
	\uX = \left\{\begin{array}{l l l}
		\left[\begin{array}{c}
			X(0)\\
			\X\\
			\X^{\ast}
		\end{array}\right], & \X \in \domC^{\frac{N-1}{2}}, & \text{if $N$ is odd}, \\[15pt]
		\left[\begin{array}{c}
			\X\\
			\X^{\ast}
		\end{array}\right], & \X \in \domC^{\frac{N}{2}}, & \text{if $N$ is even}, 
	\end{array}\right. \label{eq:uX}
\end{align}
with $\X = \{X(\omega)\}_{\omega>0}$ comprising the GDFT coefficients associated with positive frequencies. To compute the GDFT coefficients, we can thus directly evaluate 
\begin{align}
	\uX = \uF^{\Her}\x
\end{align}

\begin{figure}[ht]
	\centering
	\begin{subfigure}[t]{0.24\textwidth}
		\centering
		\includegraphics[width=0.9\textwidth, trim={4cm 11.5cm 4cm 12cm}, clip]{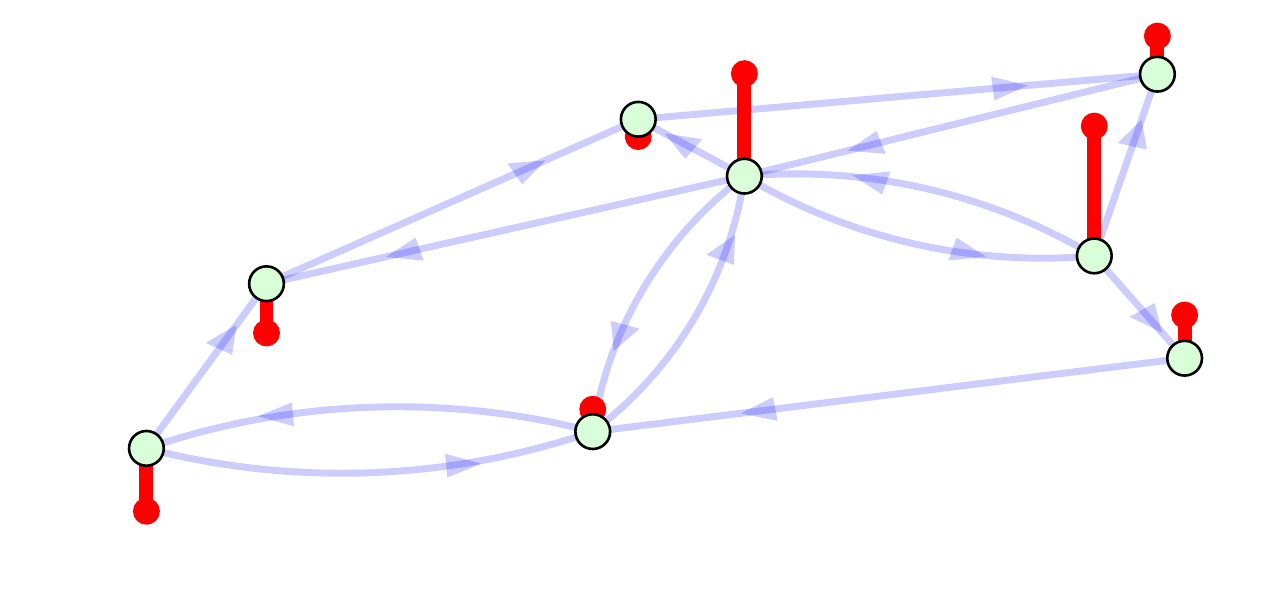} 
		\caption[]%
		{{\small $\Real{\f_{1}}$, $\omega_{1}\approx0.238$.}}    
	\end{subfigure}
	\hfill
	\begin{subfigure}[t]{0.24\textwidth}  
		\centering 
		\includegraphics[width=0.9\textwidth, trim={4cm 11.5cm 4cm 12cm}, clip]{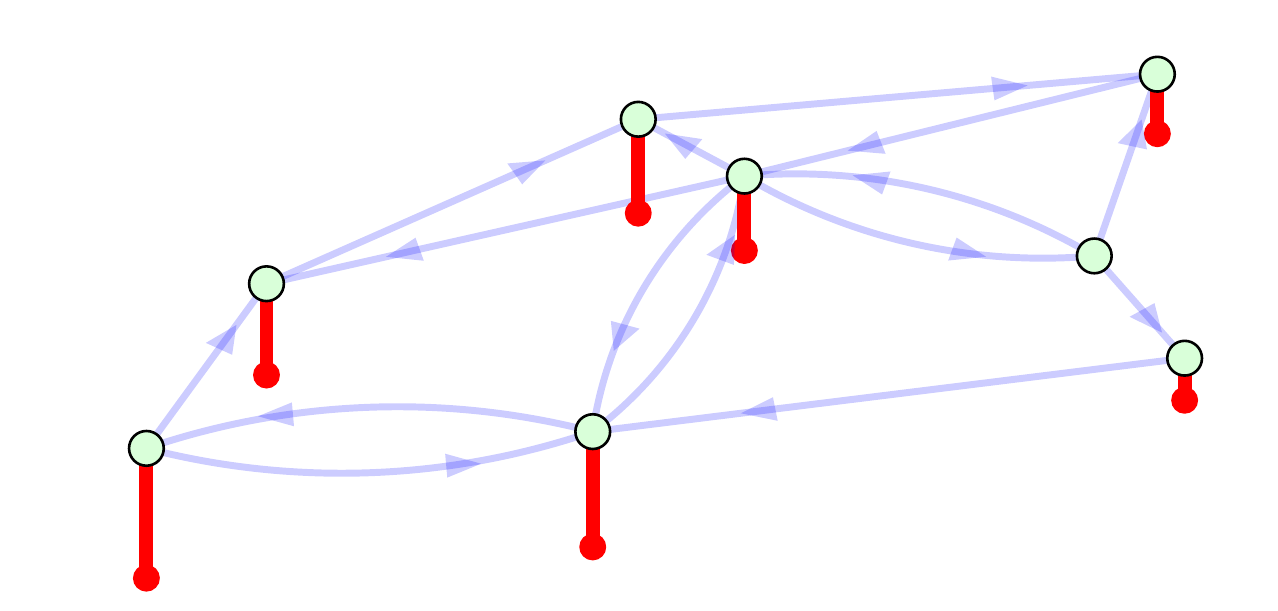} 
		\caption[]%
		{{\small $\Imag{\f_{1}}$, $\omega_{1}\approx0.238$.}}    
	\end{subfigure}
	\vspace{-0.4cm}
	\vskip\baselineskip
	\begin{subfigure}[t]{0.24\textwidth}
		\centering
		\includegraphics[width=0.9\textwidth, trim={4cm 12cm 4cm 11cm}, clip]{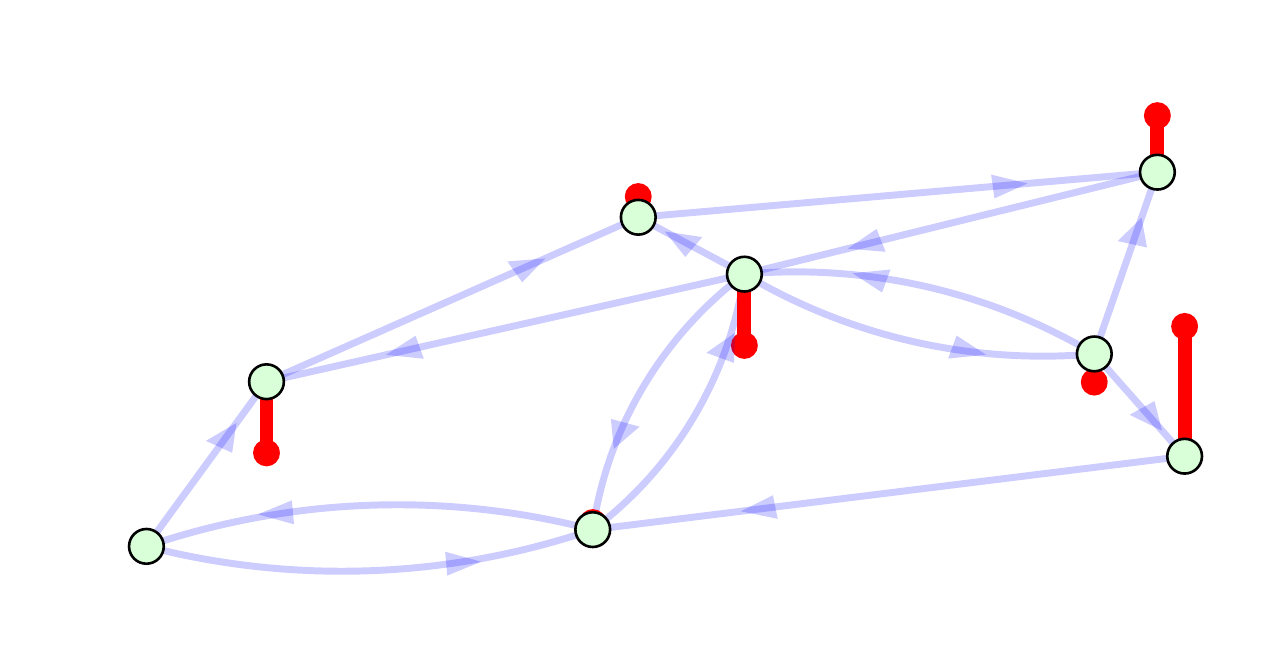} 
		\caption[]%
		{{\small $\Real{\f_{2}}$, $\omega_{2}\approx1.244$.}}    
	\end{subfigure}
	\hfill
	\begin{subfigure}[t]{0.24\textwidth}  
		\centering 
		\includegraphics[width=0.9\textwidth, trim={4cm 12cm 4cm 11cm}, clip]{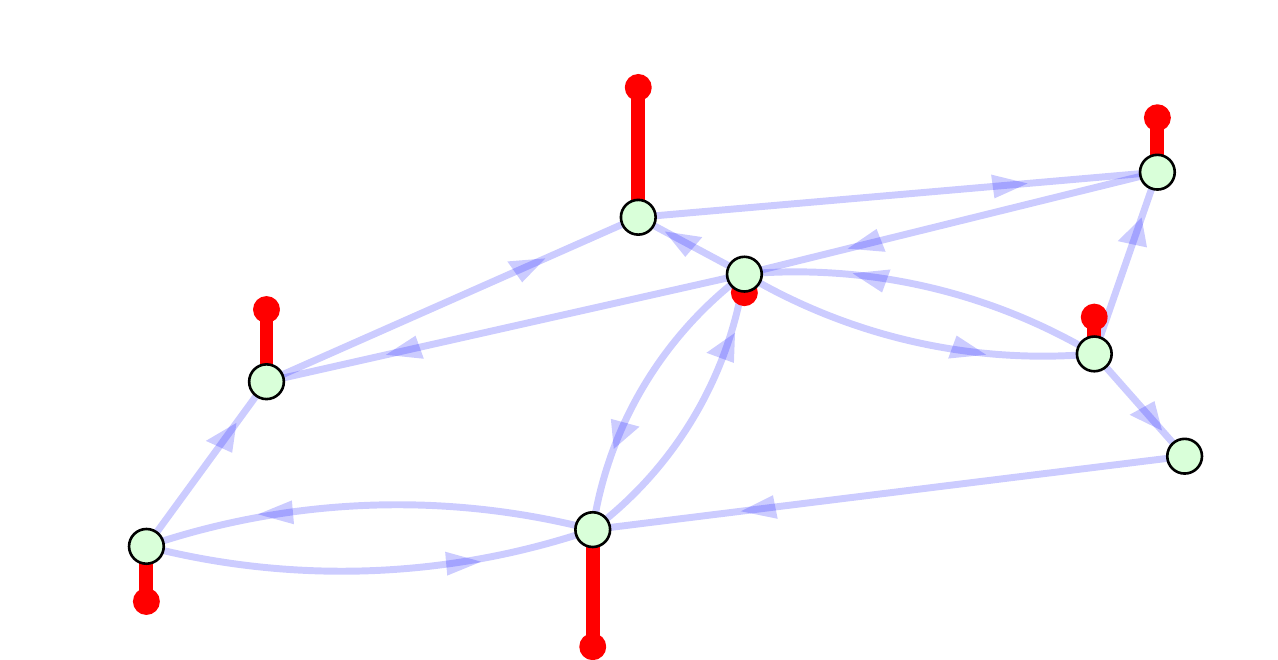} 
		\caption[]%
		{{\small $\Imag{\f_{2}}$, $\omega_{2}\approx1.244$.}}    
	\end{subfigure}
	\vspace{-0.4cm}
	\vskip\baselineskip
	\begin{subfigure}[t]{0.24\textwidth}
		\centering
		\includegraphics[width=0.9\textwidth, trim={4cm 11.5cm 4cm 11.5cm}, clip]{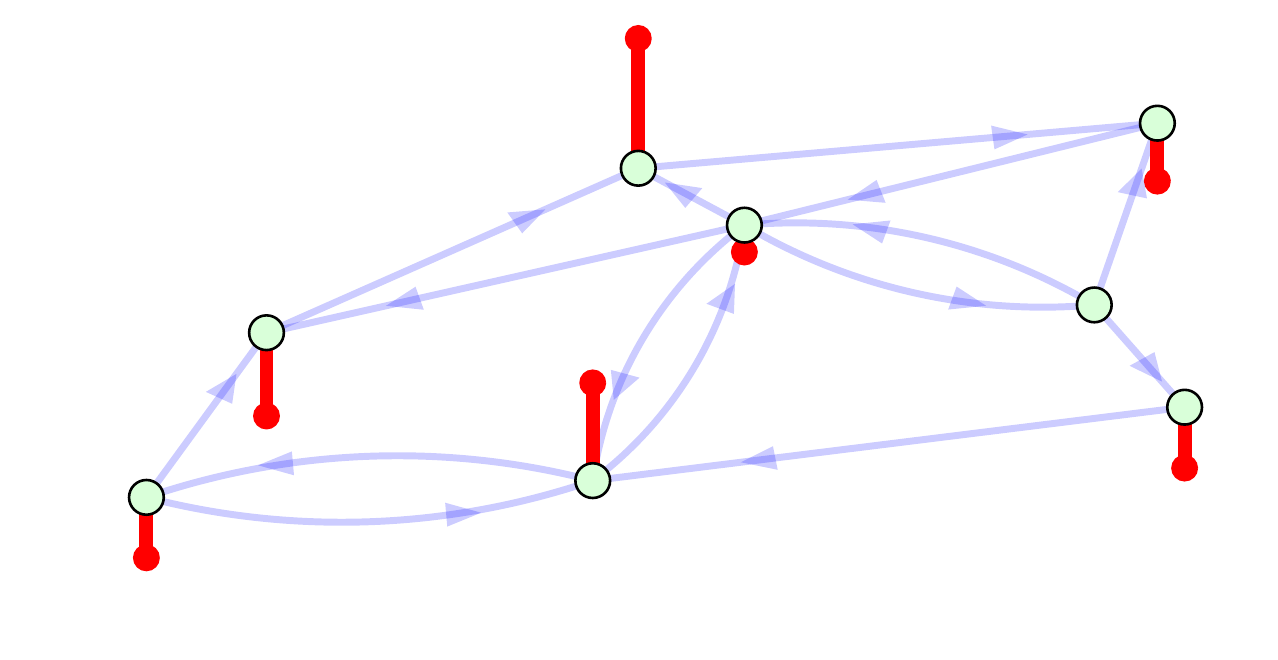} 
		\caption[]%
		{{\small $\Real{\f_{3}}$, $\omega_{3}\approx2.209$.}}    
	\end{subfigure}
	\hfill
	\begin{subfigure}[t]{0.24\textwidth}  
		\centering 
		\includegraphics[width=0.9\textwidth, trim={4cm 11.5cm 4cm 11.5cm}, clip]{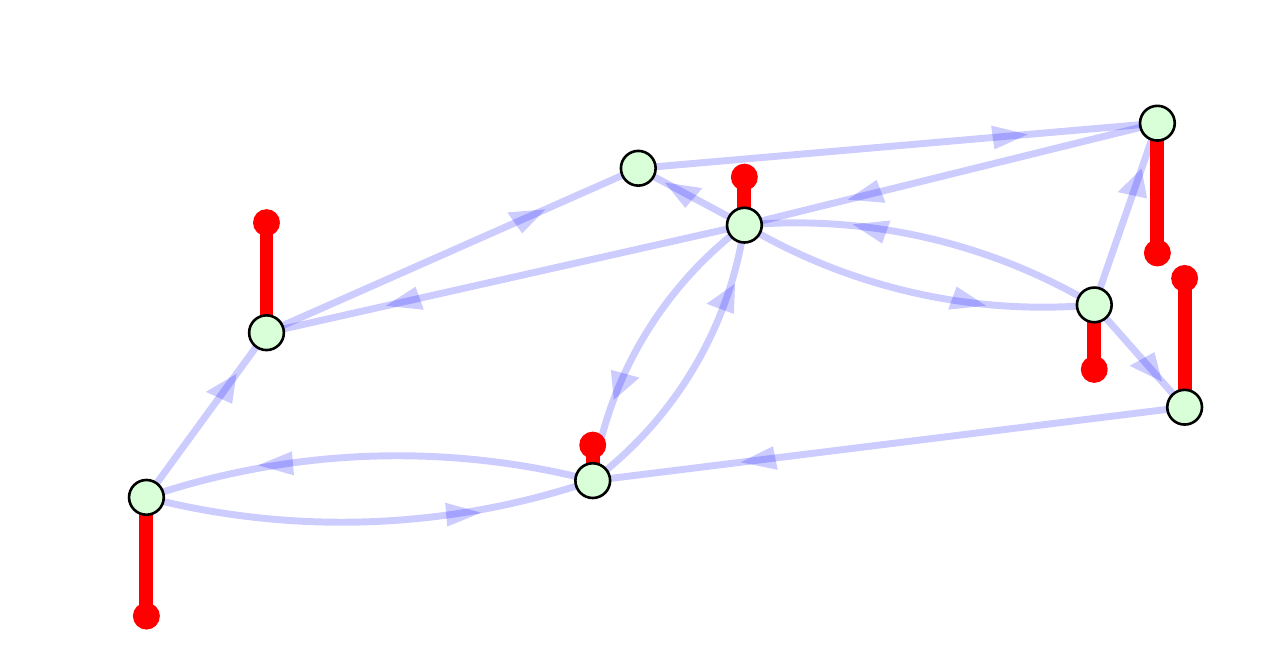} 
		\caption[]%
		{{\small $\Imag{\f_{3}}$, $\omega_{3}\approx2.209$.}}    
	\end{subfigure}
	\vspace{-0.4cm}
	\vskip\baselineskip
	\begin{subfigure}[t]{0.24\textwidth}
		\centering
		\includegraphics[width=0.9\textwidth, trim={4cm 12cm 4cm 12cm}, clip]{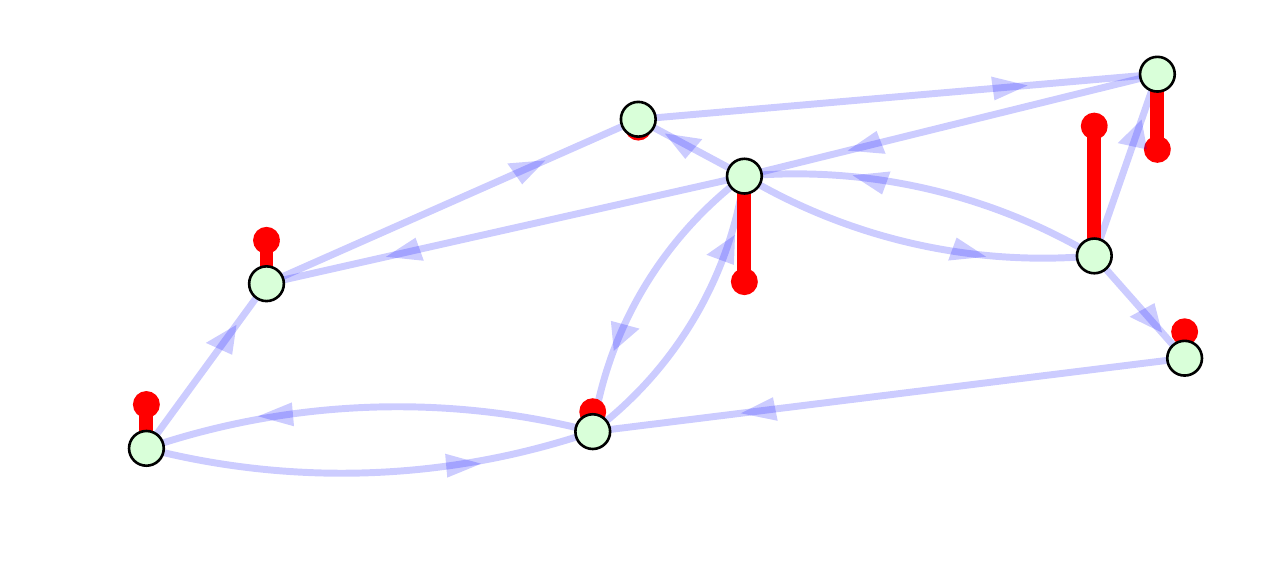} 
		\caption[]%
		{{\small $\Real{\f_{4}}$, $\omega_{4}\approx2.817$.}}    
	\end{subfigure}
	\hfill
	\begin{subfigure}[t]{0.24\textwidth}  
		\centering 
		\includegraphics[width=0.9\textwidth, trim={4cm 12cm 4cm 12cm}, clip]{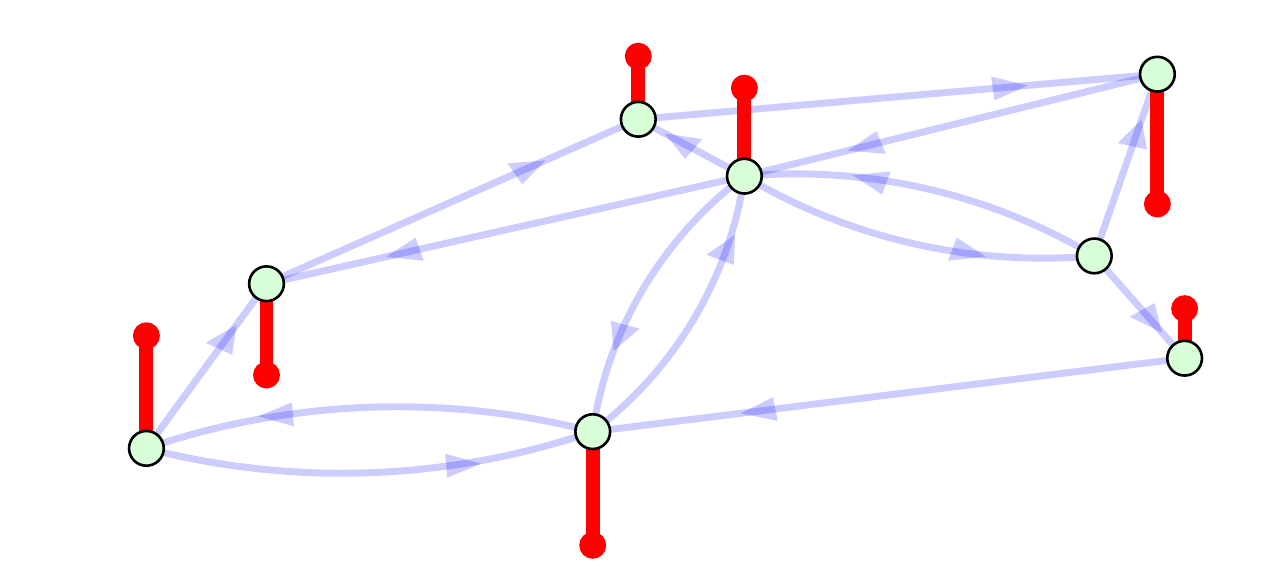} 
		\caption[]%
		{{\small $\Imag{\f_{4}}$, $\omega_{4}\approx2.817$.}}    
	\end{subfigure}
	\caption[]
	{\small The real and imaginary parts of the eigenvectors, $\f_{\omega}$, of the GSO, $\S$, of the directed graph presented in Figure \ref{fig:1}, and their corresponding angular frequencies, $\omega  \in \{0.238, 1.244, 2.209, 2.817\}$. Observe the increase in variation of $\f_{\omega}$ with the index.} 
	\label{fig:2}
\end{figure}

\begin{example}
	For the graph presented in Figure \ref{fig:structure}, the GDFT parameters, $\uF \in \domC^{8 \times 8}$ and $\uboldOmega \in \domR^{8 \times 8}$, were obtained from the eigenvalue decomposition of $\S$ as in (\ref{eq:shift_EVD}). Owing to the symmetry of the spectrum, and because here $N=8$ is an even integer, $\F \in \domC^{8 \times 4}$ contained four distinct complex-valued eigenvectors which are illustrated in Figure \ref{fig:2}. The eigenvalues in (\ref{eq:shift_EVD}), that is, the angular frequencies were given by $\omega \in \{0.238, 1.244, 2.209, 2.817\}$ in radians per second.
\end{example}

\vspace{-0.25cm}

\vspace{-0.7cm}


\subsection{System on a graph in the GDFT domain}

In light of the GDFT decomposition of $\S$ in (\ref{eq:shift_EVD}), we can now reconsider the system on a graph in (\ref{eq:system}) as follows
\begin{align}
	\y = \sum_{m=0}^{M} h_{m} \S^{m}\x = \sum_{m=0}^{M} h_{m} \, \uF e^{-\jmath m \uboldOmega} \uF^{\Her} \x
\end{align}
The pre-multiplication of this relation with $\uF^{\Her}$ yields
\begin{align}
	\uY = \sum_{m=0}^{M} h_{m} \, e^{-\jmath m \uboldOmega} \uX \label{eq:system_GDFT}
\end{align}
with the GDFT of the output signal defined as $\Y = \uF^{\Her}\y$, which exhibits a similar structure to $\uX$ in (\ref{eq:uX}). 

An element-wise formulation of the system in (\ref{eq:system_GDFT}) allows us to define the \textit{transfer function} of a system on a graph \vspace{-0.2cm}
\begin{align}
	H(\omega) = \frac{Y(\omega)}{X(\omega)} = \sum_{m=0}^{M} h_{m} e^{-\jmath \omega m}
\end{align}
which resembles the frequency response in classical Fourier setting.

\subsection{Graph discrete Hilbert transform (GDHT)}

The implicit frequency-domain interpretation of the proposed GDFT allows us to directly introduce the \textit{graph discrete Hilbert transform} (GDHT), based on the GDFT coefficients associated with the non-negative frequencies, that is
\begin{align}
	\x_{a} = \begin{cases}
		2\F\X + X(0)\f_{0}, & \text{if $N$ is odd}, \\[1pt]
		2\F\X, & \text{if $N$ is even},
	\end{cases} \label{eq:GDHT}
\end{align}
and in this way obtain the \textit{graph analytic signal}, $\x_{a} \in \domC^{N}$.

As with standard time-domain analytic signals, the proposed GSO allows for the graph analytic signal to be expressed in terms of its vertex-varying magnitude and phase, that is
\begin{align}
	x_{a}(n) = a(n)e^{\jmath \phi(n)}
\end{align}
where $a(n) = |x_{a}(n)|$ and $\phi(n) = \angle \, x_{a}(n) $ are respectively the \textit{local graph magnitude} and \textit{phase}. The \textit{local angular frequency} can also be evaluated as the differential of the local phase with respect to the vertex index, that is, $\omega(n) = \frac{d \phi(n)}{d n}$. Based on the proposed graph differential operator in (\ref{eq:graph_differential}), the local graph phase is obtained as follows
\begin{align}
	\boldomega = \boldnabla\boldphi
\end{align}

\begin{figure}[ht]
	\centering
	\begin{subfigure}[t]{0.24\textwidth}
		\centering
		\includegraphics[width=0.9\textwidth, trim={4cm 12cm 4cm 12cm}, clip]{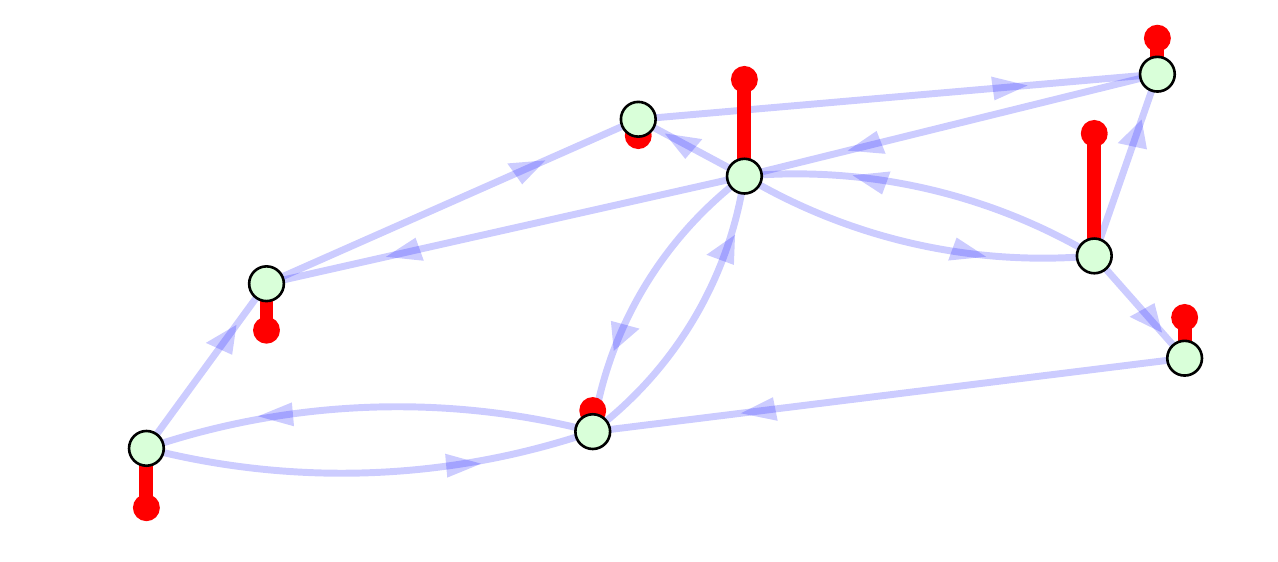} 
		\caption[]%
		{{\small Graph signal, $\x$.}}    
		\label{fig:orig2}
	\end{subfigure}
	\hfill
	\begin{subfigure}[t]{0.24\textwidth}  
		\centering 
		\includegraphics[width=0.9\textwidth, trim={4cm 12cm 4cm 12cm}, clip]{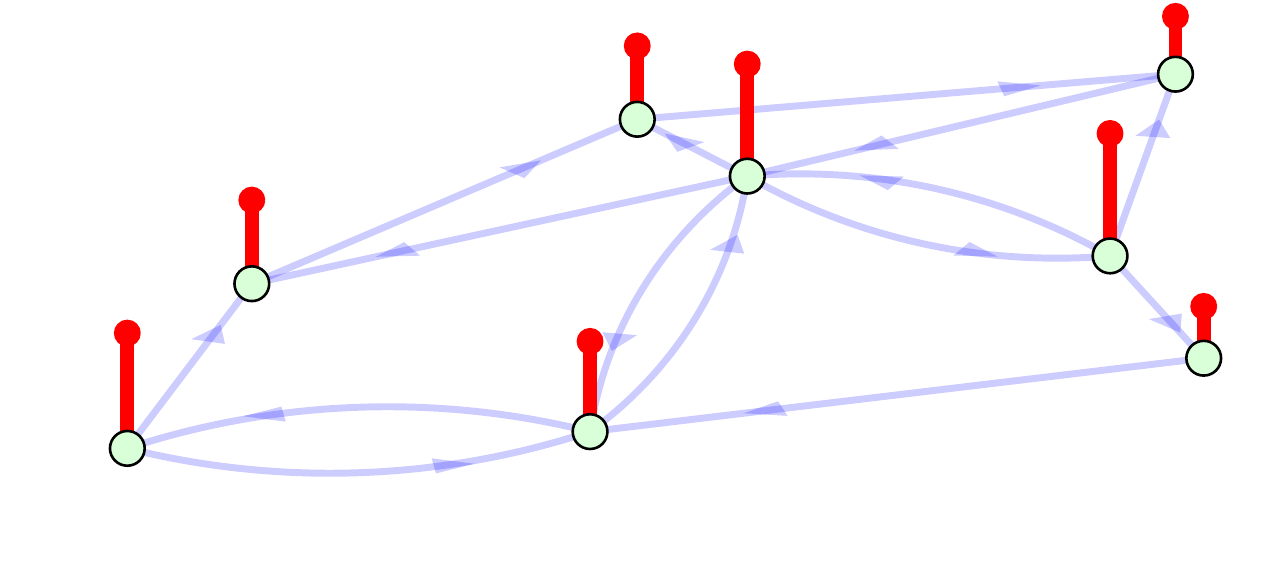} 
		\caption[]%
		{{\small Local magnitude, $\a$.}}    
		\label{fig:env}
		\end{subfigure}
	\vspace{-0.4cm}
	\vskip\baselineskip
	\begin{subfigure}[t]{0.24\textwidth}   
		\centering 
		\includegraphics[width=0.9\textwidth, trim={4cm 12.8cm 4cm 12cm}, clip]{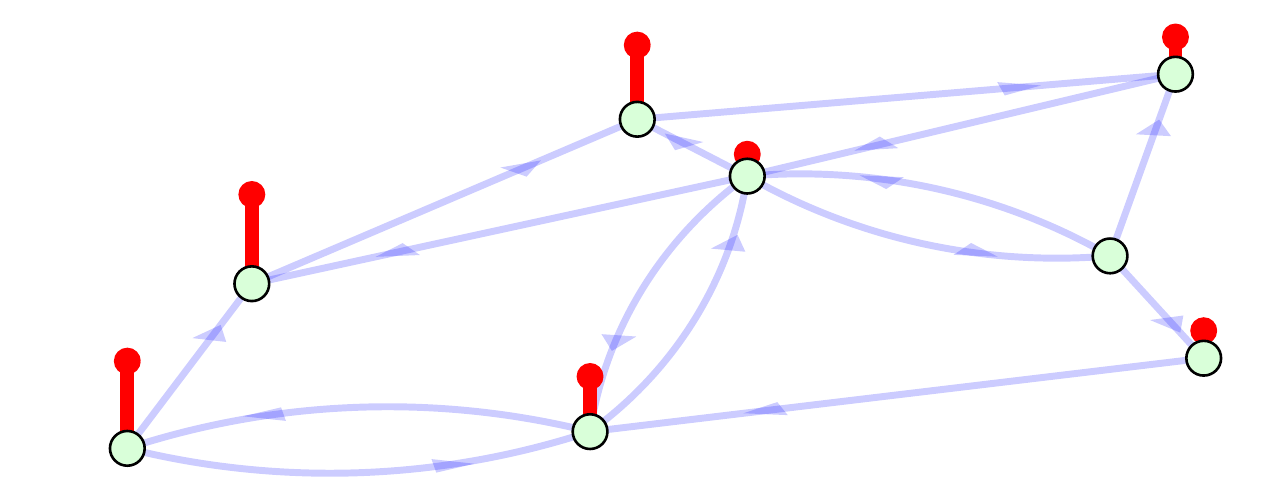} 
		\caption[]%
		{{\small Local phase, $\boldphi$.}}    
		\label{fig:phase}
	\end{subfigure}
	\hfill
	\begin{subfigure}[t]{0.24\textwidth}   
		\centering 
		\includegraphics[width=0.9\textwidth, trim={4cm 12.8cm 4cm 12cm}, clip]{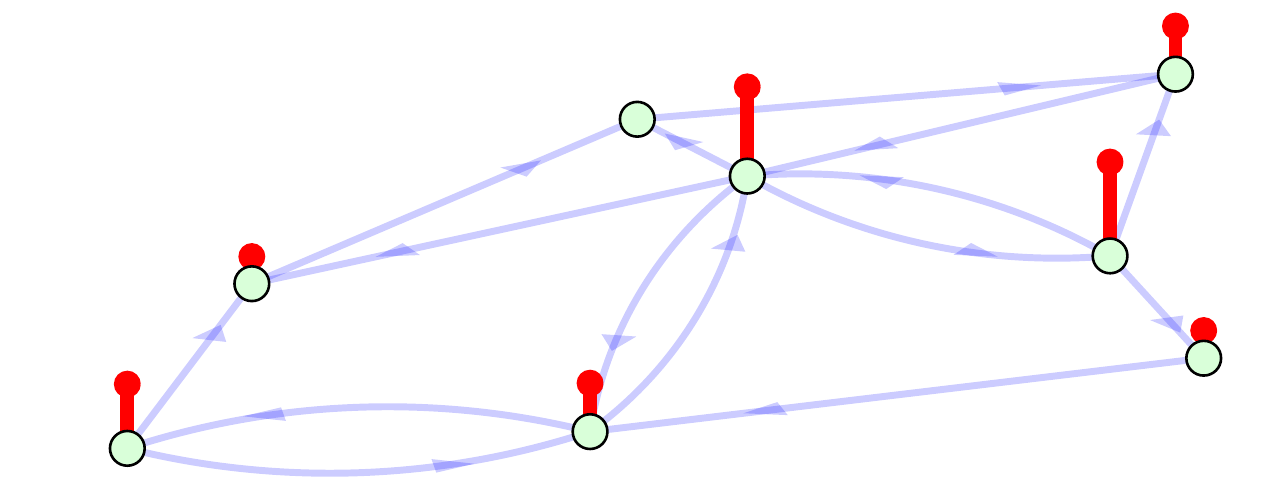} 
		\caption[]%
		{{\small Local frequency, $\boldomega$.}}    
		\label{fig:freq}
	\end{subfigure}
	\caption[]
	{\small Graph discrete Hilbert transform. (a) A signal, $\x$, on a directed graph. (b) Local graph magnitude, $\a$. (c) Local graph phase, $\boldphi$. (d) Local graph frequency, $\boldomega$.} 
	\label{fig:3}
\end{figure}

\begin{example}
	For the graph signal, $\x \in \domR^{8}$, in Figure \ref{fig:orig2}, the local graph magnitude, phase and frequency, denoted respectively by $\a, \boldphi,\boldomega \in \domR^{8}$, were evaluated using the GDHT in (\ref{eq:GDHT}) and based on the proposed GSO of the directed graph in Figure \ref{fig:structure}. Notice that the GDHT provides a means to detect features of the graph signal, such as the local magnitude, phase and frequency, which are not obvious from the signal itself.
\end{example}

\vspace{-0.5cm}

\section{Conclusions}

A class of unitary shift operators for signals on a graph has been proposed which exhibits the energy preservation property. Based on the proposed GSO, rigorous definitions of the differential operator, discrete Fourier transform and discrete Hilbert transform have been introduced and their practical utility has been demonstrated through intuitive examples.




\footnotesize
\balance

\bibliographystyle{IEEEtran}
\bibliography{Bibliography} 

\begin{thebibliography}{10}
\providecommand{\url}[1]{#1}
\csname url@samestyle\endcsname
\providecommand{\newblock}{\relax}
\providecommand{\bibinfo}[2]{#2}
\providecommand{\BIBentrySTDinterwordspacing}{\spaceskip=0pt\relax}
\providecommand{\BIBentryALTinterwordstretchfactor}{4}
\providecommand{\BIBentryALTinterwordspacing}{\spaceskip=\fontdimen2\font plus
\BIBentryALTinterwordstretchfactor\fontdimen3\font minus
  \fontdimen4\font\relax}
\providecommand{\BIBforeignlanguage}[2]{{%
\expandafter\ifx\csname l@#1\endcsname\relax
\typeout{** WARNING: IEEEtran.bst: No hyphenation pattern has been}%
\typeout{** loaded for the language `#1'. Using the pattern for}%
\typeout{** the default language instead.}%
\else
\language=\csname l@#1\endcsname
\fi
#2}}
\providecommand{\BIBdecl}{\relax}
\BIBdecl

\bibitem{Sandryhaila2013}
A.~Sandryhaila and J.~M.~F. Moura, ``{Discrete Signal Processing on Graphs},''
  \emph{{IEEE Transactions on Signal Processing}}, vol.~61, no.~7, pp.
  1644--1656, 2013.

\bibitem{Shuman2013}
D.~I. Shuman, S.~K. Narang, P.~Frossard, A.~Ortega, and P.~Vandergheynst,
  ``{The Emerging Field of Signal Processing on Graphs: Extending
  High-Dimensional Data Analysis to Networks and Other Irregular Domains},''
  \emph{{IEEE Signal Processing Magazine}}, vol.~30, pp. 83--98, 2013.

\bibitem{Ortega2018}
A.~Ortega, P.~Frossard, J.~Kova$\check{\text{c}}$evi\'c, J.~M.~F. Moura, and
  P.~Vandergheynst, ``{Graph Signal Processing: Overview, Challenges, and
  Applications},'' \emph{{In Proceedings of the IEEE}}, vol. 106, no.~5, pp.
  808--828, 2018.

\bibitem{Stankovic2019_1}
L.~Stankovi\'c, D.~P. Mandic, M.~Dakovi\'c, M.~Brajovi\'c, B.~Scalzo~Dees, and
  T.~Constantinides, ``{Graph Signal Processing -- Part I: Graphs, Graph
  Spectra, and Spectral Clustering},'' \emph{{arXiv:1907.03467 [cs.IT]}}, 2019.

\bibitem{Girault2015}
B.~Girault, P.~Goncalves, and E.~Fleury, ``{Translation on Graphs: An Isometric
  Shift Operator},'' \emph{IEEE Signal Processing Letters}, vol.~22, no.~12,
  pp. 2416--2420, 2015.

\bibitem{Girault2015_2}
B.~Girault, ``{Stationary Graph Signals using an Isometric Graph
  Translation},'' \emph{In Proceedings of the European Signal Processing
  Conference}, pp. 1516--1520, 2015.

\bibitem{Gavili2017}
A.~Gavili and X.~P. Zhang, ``{On the Shift Operator, Graph Frequency, and
  Optimal Filtering in Graph Signal Processing},'' \emph{IEEE Transactions on
  Signal Processing}, vol.~65, no.~23, pp. 6303--6318, 2017.

\bibitem{Fillmore1974}
P.~A. Fillmore, ``{The Shift Operator},'' \emph{{American Mathematical
  Monthly}}, vol.~81, no.~7, pp. 717--723, 1974.

\bibitem{Lowdin1950}
P.~O. L\"owdin, ``{On the Non‐Orthogonality Problem Connected with the Use of
  Atomic Wave Functions in the Theory of Molecules and Crystals},''
  \emph{{Journal of Chemical Physics}}, vol.~18, no.~3, pp. 365--375, 1950.

\bibitem{Lowdin1970}
------, ``{On the Nonorthogonality Problem},'' \emph{{Advances in Quantum
  Chemistry}}, vol.~5, pp. 185--199, 1970.

\bibitem{Schonemann1966}
P.~H. Sch\"onemann, ``{A Generalized Solution of the Orthogonal Proctrustes
  Problem},'' \emph{{Psychometrika}}, vol.~{31}, pp. 1--10, 1966.

\bibitem{Kabsch1976}
W.~Kabsch, ``{A Solution for the Best Rotation to Related Two Sets of
  Vectors},'' \emph{{Acta Crystallographica}}, vol.~32, p. 922, 1976.

\bibitem{Pratt1956}
G.~W. Pratt and S.~F. Neustadter, ``{Maximal Orthogonal Orbitals},''
  \emph{{Physical Review}}, vol. 101, pp. 1248--1250, 1956.

\bibitem{Stankovic2019_2}
L.~Stankovi\'c, D.~P. Mandic, M.~Dakovi\'c, I.~Kisil, E.~Sejdi\'c, and A.~G.
  Constantinides, ``{Understanding the Basis of Graph Signal Processing via an
  Intuitive Example-Driven Approach},'' \emph{{IEEE Signal Processing
  Magazine}}, in press, 2019.

\bibitem{Stone1930}
M.~H. Stone, ``{Linear Transformations in Hilbert Space: III. Operational
  Methods and Group Theory},'' \emph{{Proceedings of the National Academy of
  Sciences of the United States of America}}, vol.~16, no.~2, pp. 172--175,
  1930.

\bibitem{Stone1932}
------, ``{On One-Parameter Unitary Groups in Hilbert Space},'' \emph{{Annals
  of Mathematics}}, vol.~33, no.~3, pp. 643--648, 1932.

\end{thebibliography}

\end{document}